\documentclass{JHEP3}

\usepackage{epsfig,amssymb,amsmath,latexsym}
\usepackage{axodraw}

\newcommand\tb{\bar\theta}
\newcommand\g{\gamma}
\renewcommand\t{\theta}
\renewcommand\l{\lambda}
\renewcommand\a{\alpha}

\newcommand\GG{\xi}
\newcommand\Z{\omega}
\renewcommand\S{\Sigma}
\newcommand\p{\phi}
\renewcommand\P{\Phi}
\renewcommand\L{\ensuremath{{\cal L}}}
\newcommand\e{{\rm e}}

\newcommand\R[1]{\ensuremath{\mathbf{#1}}}
\newcommand\Rb[1]{\ensuremath{\mathbf{\bar{#1}}}}

\newcommand\F{\ensuremath{{\rm d^2\t}}}
\newcommand\D{\ensuremath{{\rm d^4\t}}}

\newcommand\mg{m_{\tilde g}}

\newcommand\ddt{\frac{\rm d}{{\rm d}t}}

\newcommand\tr{\ensuremath{\mathop{\rm tr}}}
\renewcommand\Re{\ensuremath{\mathop{\rm Re}}}

\newcommand\GeV{\mbox{GeV}}
\newcommand\TeV{\mbox{TeV}}
\newcommand\MG{M_{G}}
\newcommand\MH{M_{H}}

\newcommand\pippo[1]{\noindent\fbox{\vbox{\hsize.984\columnwidth\noindent\ignorespaces#1}}}
\newcommand\poppo{\medskip\hrule\medskip\par\noindent {\LARGE\raise-.2ex\hbox{$\bullet$}}\ }

\renewcommand\pippo[1]{}

\newcommand\be{\begin{equation}}
\newcommand\ee{\end{equation}}
\newcommand\bea{\begin{eqnarray}}
\newcommand\eea{\end{eqnarray}}
\newcommand\ba{\begin{array}}
\newcommand\ea{\end{array}}
\newcommand\bma{\begin{array}{cccc}}
\newcommand\ema{\end{array}}
\newcommand\matr[1]{\left(\bma#1\ema\right)}
\newcommand\0{\nonumber}

\title{Soft SUSY breaking contributions to proton decay}

\author{Zurab Berezhiani, Fabrizio Nesti, Luigi Pilo\\
        Dipartimento di Fisica, Universit\`a dell'Aquila\\
        67010 Coppito, L'Aquila, Italy, and\\ 
        INFN, Laboratori Nazionali del Gran Sasso\\
        67100 Assergi, L'Aquila, Italy 
}

\abstract{We show that in supersymmetric grand unified theories new
  effective D=4 and D=5 operators for proton decay are induced by soft
  SUSY-breaking terms, when heavy GUT gauge bosons are integrated out,
  in addition to the standard D=6 ones.  As a result, the proton
  lifetime in gauge mediated channels can be enhanced or even
  suppressed depending on the size of the heavy Higgses soft terms.}

\begin{document}

\section{Introduction}

%\pippo{ Intro D=5 and D=6 }

One of the immediate consequences of Grand-Unified Theories (GUT) is
baryon number non-conservation that can lead to proton
decay~\cite{gut,nathreview}. The heavy gauge bosons mediate the
effective baryon-number violating four-fermion operators
\begin{alignat}{4}
\label{eq:opG}
&O_{\rm gauge} \qquad&\sim &\qquad g_G^2\MG^{-2}\,(\overline q\,u^c \,\overline l \,d^c)\,,&\qquad  &
g^2_G\MG^{-2}\,(\overline q\, u^c\, \overline q\,  e^c)\,,
\end{alignat}
where $\MG$ and $g_G$ are the Grand Unification scale and gauge
coupling constant, $q=(u,d)_L$, $l=(\nu,e)_L$ are the left-handed
quarks and leptons (weak isodoublets) and $u^c$, $d^c$, $e^c$ are
charge-conjugated fields of the right handed ones: $u_R$, $d_R$, $e_R$
(weak isosinglets). In addition, other four-fermion operators of
different structure are mediated by heavy colored triplet Higgses with
mass $\MH\sim\MG$:
\begin{alignat}{4}
\label{eq:opH}
&O_{\rm higgs} \qquad&\sim &\qquad g_Y^2\MH^{-2}(q\, q\, q\, l)\,,    &\qquad  & g^2_Y\MH^{-2}(u^c\, u^c\, d^c\, e^c)\,,
\end{alignat}
The latter operators are typically weaker because of smallness of the
Yukawa couplings $g_Y$ but in some models they can be dominant over
the gauge mediated operators (\ref{eq:opG})~\cite{john}. 

The set~(\ref{eq:opG}), (\ref{eq:opH}) represents all possible D=6
baryon number violating operators, independently of the details of
grand unification~\cite{W}. The two kinds of operators have different
chirality structures (LRLR for~(\ref{eq:opG}) and LLLL or RRRR
for~(\ref{eq:opH})) and they could in principle be distinguished via
the polarization of final states~\cite{W}. Both kinds are suppressed by
two powers of $\MG$.

In non supersymmetric models, when $\MG$ is below $10^{15}\,\GeV$,
processes following from~(\ref{eq:opG}) are already ruled
out. Supersymmetry, in addition to making the unification more
natural, raises its scale, setting the magnitude of these processes
within the reach of near future experimental facilities.

Supersymmetry raises the unification scale to $10^{16}\,\GeV$ and
makes these operators hardly observable. On the other hand it
introduces additional D=5 baryon number violating
operators mediated by heavy colored higgsinos~\cite{dimfive}, as
\be
O_{\rm higgsino} \qquad\sim\qquad g_Y^2\MH^{-1}(\tilde q\,  q\, \tilde q\, l)\,,\qquad
g_Y^2\MH^{-1}(\tilde u^c\, u^c\, \tilde d^c\, e^c)\,,
\ee
where tilded fields represent scalar superpartners. These are
suppressed by a single power of the GUT scale and after dressing by
gaugino exchange they give rise to operators of the
form~(\ref{eq:opH}) with a cutoff scale $\sim (M_H m_{S})^{1/2}$ where
$m_S$ is the SUSY-breaking scale~\cite{dimfive}. They thus become
generically dominant and on the verge of being in conflict with
current experimental limits on these specific decay modes ($p\to K\nu$
etc.)~\cite{nathreview}.  However, their magnitude is very model
dependent: essentially they exclude minimal versions GUTs and cause
problems for models unless fine tuning is arranged.  Several
mechanisms have been devised to suppress them by playing with the
structure of the heavy sector of the theories.\footnote{There are
  several ideas how dimension-5 operators can be suppressed. In
  particular this can be due to special arrangements in the heavy
  higgs sector~\cite{bbv}, because of symmetry properties of the
  Yukawa sector~\cite{bvt}. In SO(10) models, LLLL operators can be
  naturally suppressed by the choice of the SO(10) breaking VEVs while
  the less dangerous RRRR ones are left allowed. With further model
  building also these latter can be eliminated~\cite{gia}. Finally, in
  supersymmetry there are also D=4 B and L violating operators that
  can be forbidden by exact R-parity~\cite{Rparity}.}

\pippo{ Fake supersymmetry gives D=4 operators.}

Let us remark that gauge coupling unification does not strictly
require supersymmetry of the theory. For instance the presence of
fermionic partners of the gauge and higgs at $\TeV$ scale can adjust
the running of the gauge coupling constants so that they unify at one
point.  Though scalars are not crucial for unification they are
predicted by low scale SUSY; however, finding at LHC a SUSY-like
spectrum would not mean that supersymmetry is discovered. Indeed, it
would be extremely difficult to verify that the lagrangian has a
supersymmetric structure, i.e.\ that the different coupling
constants are related, like the quark-squark-gluino coupling constants
that should be exactly the same as the strong gauge coupling constant.

One can imagine a \emph{fake-SUSY} theory where only the sparticle
\emph{spectrum} is supersymmetric (i.e.\ every particle has its
``superpartner'') while the the lagrangian is not.  What would happen
in such a theory?  We argue that, even if the gauge coupling
unification is achieved as perfectly as in a truly supersymmetric
theory, it would lead to disastrous proton decay rate.  The reason is
the following: once such a theory contains scalars partners of quarks
and leptons ($\tilde q$, $\tilde l$) it generically contains D=4
operators of the form
\begin{alignat}{4}
\label{eq:opQ}
&O_{\rm quartic} \qquad&\sim &\qquad (\tilde q^*\,\tilde u^c \,\tilde l^* \,\tilde d^c)\,,&\qquad  &
(\tilde q^*\, u^c\, \tilde q^*\,  e^c)\,,
\end{alignat}
which in a GUT context can not be excluded by any symmetry
reason. Notice that even if they are not present in the bare
lagrangian they emerge radiatively by loops of GUT gauge bosons.  The
dressing by gauginos transforms these D=4 into D=6 ones on the
form~(\ref{eq:opG}) that directly cause the proton to decay at a
dangerous high rate, being suppressed only by two powers of the fake
superpartners mass scale that is of order $\TeV$.

\pippo{ Real susy cancels: Example of U(1) and general.}

Complete supersymmetry instead provides an automatic protection from
these D=4 operators: they in fact correspond to the D-terms relative
to the broken gauge generators, and they have to vanish if SUSY is
unbroken.  What happens is that the existing D-term involving light
fields $g^2|\phi^*\phi|^2$ is cancelled by two other diagrams: one
with the exchange of the (broken) gauge field and one with the
exchange of the heavy longitudinal part of the higgs field that breaks
the gauge group. For this cancellation to hold it is crucial that the
coupling $g$ is exactly the same in the three graphs, i.e.\ that SUSY
is exact.  The shortest proof of this fact can be given in the
superfield formalism, where the only supersymmetric D-term involving
four light superfields is $[\P^\dagger\P\P^\dagger\P]_D$. This
operator does not contain a four scalar contact interaction, that
therefore has to vanish.

\pippo{Real case and missing analysis}

However, since supersymmetry has to be broken, one expects that this
protection mechanism works only partially and that the susy-breaking
terms will turn on such operators. As a result they may significantly
affect the proton decay. Indeed, it was noted in~\cite{DS} that
SUSY-breaking induces the D=4 scalar operators
(\ref{eq:opQ}). However, surprisingly enough a complete analysis of
the soft-susy breaking effects on proton decay has not been
performed.\footnote{In~\cite{sak1} it is described a classification of
  all the D=4, 5, 6 operators relevant for proton decay, while in
  \cite{sak2} the effect of soft terms in a SUGRA scenario was
  studied, but only for the analytic D=5, 4 operators.}

\medskip

\pippo{Content}

In this work we study the effect of the soft terms on the low energy
effective theory produced after the heavy gauge superfields are
integrated out at the GUT scale.  We show that the D=6 operators are
always accompanied by new operators of D=5 and D=4, turned on by the
presence of the soft terms.  Next, we compute the renormalization of
these operators from the GUT scale to the SUSY breaking scale; we
adopt the techniques illustrated in~\cite{kazakov} that simplify
considerably the task.  We then dress the new D=5 and D=4 operators at
the SUSY breaking scale, transforming them in the form~(\ref{eq:opG})
and estimate when they can be relevant. As an example, we discuss the
SUSY SU(5) model and show that their contribution can be important and
could bring the proton decay rate in specific channels to be
experimentally accessible.

\section{Gauge mediated effective operators in softly broken SUSY GUT}

\pippo{ Soft decoupling gives D=5+4 and D=6+5+4. }

The gauge mediated effective operators are efficiently described in
the superfield formalism with soft breaking terms inserted as
spurions.  If we arrange the chiral superfields of irreducible
representations in a column vector $\P=\{\P_I\}$, the full lagrangian
is, in compact notation:
\be
\label{eq:L}
\L=\int \!\D\, \Big [ \P^\dagger X \e^{2gV}\P\Big ]
  +\int \!\F\, \Big [W(\P)+W_\a W^\a Y \Big]+h.c.\,.
\ee
Here the gaugino masses enter via $Y=(1+\mg\t^2)$, and the soft
D-terms may be parametrized by the matrix $X=(1 +\Gamma\t^2
+\Gamma^\dagger\tb^2 +Z\t^2\tb^2)$ via the matrices $\{\Gamma_{IJ}\}$
of order $m_S$ and $\{Z_{IJ}\}$ of order $m_S^2$. One can however
perform a field redefinition to set $\Gamma_{IJ}=0$. For simplicity we
will also consider only universal soft terms, taking $Z_{IJ}$
diagonal.  Therefore we have:
\be 
X_{IJ}=X_I\,\delta_{IJ}=(1-m^2_I\t^2\tb^2) \,\delta_{IJ}\,.
\ee 
The superpotential $W(\P)$ includes the soft F-terms via spurion
fields and may be parame\-trized similarly (see e.g.~(\ref{eq:WS}))
% \footnote{Si puo` scrivere ma qui non serve: The superpotential is
%   $W(\P)=M\P^2(1+B\t^2)+\l\P^3(1+A\t^2)$, with matrix couplings
%   $M=\{M_{IJ}\}$, $\l=\{\l_{IJK}\}$ and soft terms
%   $A=\{A_{IJK}\}\sim m_S$, $B=\{B_{IJ}\}\sim m_S$.}
but its explicit form is not directly relevant for this section.

At the scale of gauge symmetry breaking one can decompose $\P_I$ in
($\P_H$, $\P_A$, $\p_i$), respectively heavy superfields, light
goldstone superfields, and light non-goldstone superfields.
The decoupling of heavy superfields $\P_H$ (e.g.\ colored higgses)
leads to  dimension-6 and analytic dimension-5 effective operators
that may violate baryon number.  The decoupling of heavy gauge fields
and goldstones in turn leads to the D-term effective operators of
dimension~6 \cite{dimfive}. All these operators are affected by the
soft susy breaking terms.

To find the effect on the gauge mediated dimension-6 operators, it is
convenient to adopt the so called super-unitary gauge~\cite{fayet},
where the goldstone superfields $\P_A$ are gauged away inside the
broken massive gauge superfields, denoted as $V_A$. To integrate out
$V_A$ one expands the gauge exponential in (\ref{eq:L}) to the quadratic
order
\bea
\label{eq:quad}
\L_{(2)}&=&\int\!\D\Big[\P^\dagger X \P +2 J_A V_A + K_{AB} V_A V_B
  + \cdots\Big]\0\\
J_A&=&g\, \P^\dagger X T_A \P\,,\0\\
K_{AB}&=&2g^2 \P^\dagger X T_AT_B\P
\eea
and then one notes that in the unitary gauge 
$$
J_A=g\,\p_i^\dagger  T_A \p_i\,X_{i}\,\qquad 
K_{AB}=2 g^2\langle\P_H\rangle^\dagger T_AT_B \langle\P_H\rangle\,X_{H}\,.
$$
As expected a VEV of the heavy fields give a squared-mass matrix
$K_{AB}$ to the broken gauge fields.\footnote{In the presence of
non-universal soft terms $J_A$ has an additional piece
$\langle\P_H\rangle^\dagger T_A X_{Hj} \p_j$, that gives new soft
masses to $\p_j$.  Also $K_{AB}$ is modified in a similar fashion,
see~\cite{decoupling}. However the present analysis is not affected
substantially.}  In a suitable basis of broken generators this matrix
is diagonal, $K_{AB}=K_A\delta_{AB}$.
We can note already at this stage that the heavy gauge boson mass
matrix contains SUSY-breaking factors, such as the $X$'s.

The result after integrating out the broken gauge fields $V_A$ is
then:
\bea
\lefteqn{\int\!\D\,J_A K^{-1}_{AB} J_B =}\0\\
&=&\!\!\int\!\D\,\frac{g^2}{2g^2
  \langle\P_H\rangle^\dagger T_AT_A
  \langle\P_H\rangle} \frac {X_i X_j}{X_H} \Big(\p_i^\dagger T_A \p_i\Big)\Big( \p_j^\dagger T_A
\p_j \Big)\,,
\label{eq:opgen}
\eea
where summation on all indices is understood. Note that in the
integration we have ignored sub-leading terms like the gauge kinetic
term and the gaugino masses for the $V_A$ gauge fields. In fact at
they both give subleading effects in~(\ref{eq:opgen}).

Considering that $\langle\P_H\rangle\sim \MG$, we recognize in the
first factor the standard coupling constant of the dimension-6
operators $\sim 1/\MG^2$.  The supersymmetry breaking however has
propagated in this operator, and indeed the second factor involves the
soft SUSY breaking D-terms $X$ that are carried along in the
decoupling process.  Moreover, due to the soft SUSY breaking in the
superpotential, also $\langle\P_H\rangle$ has in general a
non-vanishing F-term, $\langle\P_H\rangle=v_H(1+f_H\t^2)$, that
induces an other supersymmetry breaking in the effective gauge bosons
mass.  

The overall result with SUSY-breaking terms can be conveniently
rewritten as:
\be
\label{eq:operator}
\sum_{ij,A}\int\l_6 \Big(1+\GG\t^2+\GG^\dagger\tb^2+{\Z}_{ij}\t^2\tb^2\Big)\,
\Big(\p_i^\dagger T_A \p_i\Big)\Big( \p_j^\dagger T_A\p_j \Big)\,\D\,.
\ee
where $\l_6=g^2/M_A^2$ is the supersymmetric four-fermion coupling,
the heavy gauge-bosons masses are given by $M_A^2=\sum_H
2g^2(v_H^\dagger T_AT_Av_H)$, and the SUSY-breaking coefficients are:
\be
\label{eq:effectivesoftterms} 
\GG=-f_H\,,   \qquad   {\Z}_{ij}=-m^2_i-m^2_j+m^2_H+|f_H|^2\,.
\ee
For simplicity in this last expression we have assumed the breaking by
a single VEV.\footnote{With more VEVs, in the first formula $f_H$
should be replaced by its ``average'' $(\sum_H M_{A\,(H)}^2\,f_H)/
(\sum_H M_{A\,(H)}^2)$, where $M^2_{A\,(H)}=2g^2v_H^\dagger
T_AT_Av_H$.  Similarly in the second formula for $m^2_H$ and
$|f_H|^2$.}

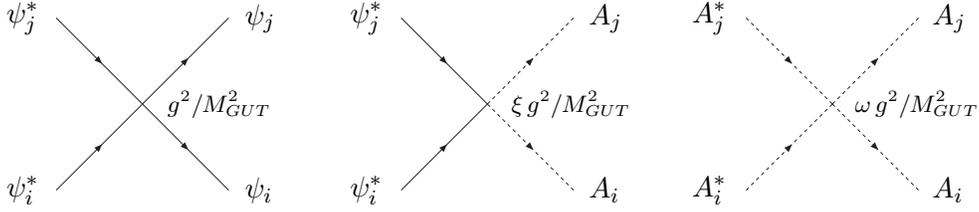
\begin{figure*}
    \newcommand\clap[1]{\hbox to 0pt{\hss#1\hss}}
    \newcommand\cclap[1]{\vbox to 0pt{\vss\hbox to 0pt{\hss#1\hss}\vss}}
\centerline{
%  \fbox{
\SetScale{.65}
\setlength\unitlength{.65\unitlength}
\begin{picture}(510,120)(0,0)
\ArrowLine(0,100)(50,50)  \ArrowLine(50,50)(100,100)
\ArrowLine(0,0)(50,50)    \ArrowLine(50,50)(100,0)
\Text(-10,0)[]{\llap{$\psi_i^*$}}
%\Vertex(50,50){4}
\Text(-10,100)[]{\llap{$\psi_j^*$}}
\Text(65,50)[]{\rlap{\footnotesize $g^2/M^2_{GUT}$}}
\Text(110,0)[]{\rlap{$\psi_i$}}
\Text(110,100)[]{\rlap{$\psi_j$}}
\ArrowLine(200,100)(250,50)  \DashArrowLine(250,50)(300,100){2}
\ArrowLine(200,0)(250,50)    \DashArrowLine(250,50)(300,0){2}
\Text(190,0)[]{\llap{$\psi_i^*$}}
\Text(190,100)[]{\llap{$\psi_j^*$}}
%\Vertex(250,50){4}
\Text(265,50)[]{\rlap{\footnotesize $\GG\,g^2/M^2_{GUT}$}}
\Text(310,0)[]{\rlap{$A_i$}}
\Text(310,100)[]{\rlap{$A_j$}}
\DashArrowLine(400,100)(450,50){2}  \DashArrowLine(450,50)(500,100){2}
\DashArrowLine(400,0)(450,50){2}    \DashArrowLine(450,50)(500,0){2}
\Text(390,0)[]{\llap{$A_i^*$}} 
\Text(390,100)[]{\llap{$A_j^*$}} 
%\Vertex(450,50){4}
\Text(465,50)[]{\rlap{\footnotesize ${\Z}\,g^2/M^2_{GUT}$}}
\Text(510,0)[]{\rlap{$A_i$}} 
\Text(510,100)[]{\rlap{$A_j$}} 
\end{picture}%
%    \setlength\unitlength{6em}
%   \begin{picture}(5.5,1.5)(-.25,-0.2)
% %    \put(0,0){\line(2,1){.5}}
% %    \put(1,0){\line(-2,1){.5}}
% %    \multiput(.5,.25)(0,.2){4}{\line(0,1){.1}}
% %    \put(0,1.2){\line(2,-1){.5}}
% %    \put(1,1.2){\line(-2,-1){.5}}
% %    \put(1.5,0.6){$\rightarrow$}
%     \put(0,0){\line(1,1){1}}
%     \put(1,0){\line(-1,1){1}}
%     \put(-0.05,0){\llap{$\psi_i^*$}}
%     \put(1.05,0){\rlap{$\psi_i$}}
%     \put(-0.05,.95){\llap{$\psi_j^*$}}
%     \put(1.05,.95){\rlap{$\psi_j$}}
%     \put(0.65,0.5){\rlap{$g^2/M^2_{GUT}$}}
% %
%     \put(2.5,.5){\line(-1,1){.5}}
%     \put(2.5,.5){\line(-1,-1){.5}}
%     \multiput(2.5,.5)(.25,.25){2}{\line(1,1){.2}}
%     \multiput(2.5,.5)(.25,-.25){2}{\line(1,-1){.2}}
%     \put(1.95,0){\llap{$\psi_i^*$}}
%     \put(3.05,0){\rlap{$A_i$}}
%     \put(1.95,.95){\llap{$\psi_j^*$}}
%     \put(3.05,.95){\rlap{$A_j$}}
%     \put(2.65,0.5){\rlap{$g^2\GG/M^2_{GUT}$}}
% %
%     \multiput(4.5,.5)(.25,.25){2}{\line(1,1){.2}}
%     \multiput(4.5,.5)(-.25,.25){2}{\line(-1,1){.2}}
%     \multiput(4.5,.5)(.25,-.25){2}{\line(1,-1){.2}}
%     \multiput(4.5,.5)(-.25,-.25){2}{\line(-1,-1){.2}}
%     \put(3.95,0){\llap{$A_i^*$}}
%     \put(5.05,0){\rlap{$A_i$}}
%     \put(3.95,.95){\llap{$A_j^*$}}
%     \put(5.05,.95){\rlap{$A_j$}}
%     \put(4.65,0.5){\rlap{$g^2{\Z}/M^2_{GUT}$}}
% %
% %    \put(2,0){\line(2,1){.5}}
% %    \put(3,0){\line(-2,1){.5}}
% %    \multiput(2.5,.25)(0,.2){4}{\line(0,1){.1}}
% %    \put(2,1.2){\line(2,-1){.5}}
% %    \put(3,1.2){\line(-2,-1){.5}}
%   \end{picture}
}
  \caption{New operators generated by soft SUSY breaking terms in the
  heavy gauge exchange.}
  \label{fig:operators}
\end{figure*}

In terms of field components the effective operator
(\ref{eq:operator}) contains the three operators shown in
figure~\ref{fig:operators}: the standard dimension-6 four-fermion
operator $\psi^*\psi \psi^*\psi$ with coupling $\sim1/\MG^2$; then a
new dimension-5 operator of the form $A^*\psi A^*\psi+h.c.$ with
coupling $\sim m_S/\MG^2 $ coming from the terms with $\t^2$ and
$\tb^2$, and finally a new dimension-4 operator of the form
$AA^*AA^*$, with coupling $\sim m_S^2/\MG^2$.

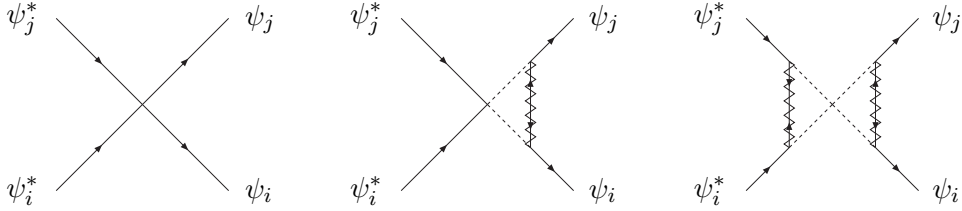
\begin{figure*}
\centerline{
%  \fbox{
\SetScale{.65}
\setlength\unitlength{.65\unitlength}
\begin{picture}(510,120)(0,0)
\ArrowLine(0,100)(50,50)  \ArrowLine(50,50)(100,100)
\ArrowLine(0,0)(50,50)    \ArrowLine(50,50)(100,0)
\Text(-10,0)[]{\llap{$\psi_i^*$}}
\Text(-10,100)[]{\llap{$\psi_j^*$}}
\Text(110,0)[]{\rlap{$\psi_i$}}
\Text(110,100)[]{\rlap{$\psi_j$}} 
\ArrowLine(200,100)(250,50)  \DashLine(250,50)(275,75){2} \ArrowLine(275,75)(300,100)  
\ArrowLine(200,0)(250,50)    \DashLine(250,50)(275,25){2} \ArrowLine(275,25)(300,0)    
\ZigZag(275,25)(275,75){3}{6}
\ArrowLine(275,50)(275,75)
\ArrowLine(275,50)(275,25)
\Text(190,0)[]{\llap{$\psi_i^*$}} 
\Text(190,100)[]{\llap{$\psi_j^*$}} 
\Text(310,0)[]{\rlap{$\psi_i$}} 
\Text(310,100)[]{\rlap{$\psi_j$}} 
\ArrowLine(400,100)(425,75) \DashLine(425,75)(450,50){2}     \DashLine(450,50)(475,75){2} \ArrowLine(475,75)(500,100)    
\ArrowLine(400,0)(425,25)   \DashLine(425,25)(450,50){2}       \DashLine(450,50)(475,25){2} \ArrowLine(475,25)(500,0)      
\ZigZag(425,25)(425,75){3}{6}
\ArrowLine(425,75)(425,50)
\ArrowLine(425,25)(425,50)
\ZigZag(475,25)(475,75){3}{6}
\ArrowLine(475,50)(475,75)
\ArrowLine(475,50)(475,25)
\Text(390,0)[]{\llap{$\psi_i^*$}} 
\Text(390,100)[]{\llap{$\psi_j^*$}} 
\Text(510,0)[]{\rlap{$\psi_i$}} 
\Text(510,100)[]{\rlap{$\psi_j$}} 
\end{picture}%
}
  \caption{four-fermion operators after dressing via gluino exchange.}
  \label{fig:dressing}
\end{figure*}

The new dimension-5 and dimension-4 operators can be dressed by
gaugino exchange at the SUSY-breaking scale (see
figure~\ref{fig:dressing}) and transformed in effective dimension-6
four-fermion operators, as it happens for dimension-5 analytic
operators.  Each dressing loop brings a factor $\sim 1/m_S$, so that
the effective strength of all these operators is the same,
$1/\MG^2$. The actual relative strength will depend on the coupling
constants involved in the dressing and on the ratio of the effective
soft breaking parameters $\GG$, ${\Z}$ to the gaugino and/or sfermion
masses.

\section{Running and dressing}

To calculate the effect of the three operators in~(\ref{eq:operator})
one has to run them from the decoupling (GUT) scale down to the SUSY
breaking scale and dress them to get the dimension-6 effective
operators. The running below the SUSY scale is non-supersymmetric and
was analyzed in~\cite{ellis}.

\medskip

\pippo{ Running from GUT }

Renormalization mixes the supersymmetric and the non-supersymmetric
effective operators via the soft susy breaking parameters of the
theory, mainly the gaugino masses. The detailed computation is rather
complicated due to the large number of diagrams involved.

Instead of attacking the problem by brute force, we employ the elegant
techniques devised in~\cite{kazakov} to analyze the soft terms
renormalization.  Starting from the anomalous dimensions of the
supersymmetric operators, we find the renormalization in the softly
broken theory by promoting the couplings to full superfields built
with the soft terms.

\pagebreak[3]

We start from the definition of the renormalization of the
supersymmetric coupling in~(\ref{eq:operator}):
\be
\frac{\l_6^B}{\l_6}=Z_6(\a_3,\a_2,\a_1)\,,\qquad Z_6=\prod_{i=1,2,3}\!\left(\frac{\a_i^B}{\a_i}\right)^{\!\frac{-\g_6^{(i)}}{b^{(i)}}}
\ee
where $b^{(i)}$ is the beta-function coefficient for each gauge group
and $\g^{(i)}_6$ is the corresponding supersymmetric contribution to
the anomalous dimension of $\l_6$.\footnote{In the one-loop
  approximation only the renormalization due to gauge loops needs to
  be taken into account, and in the operators involving the first
  generation the contribution of large top Yukawa is suppressed by
  mixings.  In higher loop orders one should also include other
  effects, for example the threshold corrections due to insertions of
  more than one $\l_6$.}  These anomalous dimensions were calculated
in~\cite{ibanez}: $\g_6^{(3)}=-4/3$, $\g_6^{(2)}=-3/2$,
$\g_6^{(1)}\simeq-23/30$.

The key step, to renormalize the full operator
$\l_6(1+\GG\t^2+\GG^\dagger\tb^2+{\Z}\tb^2\tb^2)$ in the presence of soft terms,
is to promote each gauge coupling  $\a$ to $\tilde \a = \a
(1+\mg\t^2+\mg^*\tb^2+2|\mg|^2\t^2\tb^2)$:
\be
\frac{\l_6^B(1+\GG^B\t^2+\GG^{\dagger B}+{\Z}^B\tb^2\tb^2)}{\l_6(1+\GG\t^2+\GG^\dagger\tb^2+{\Z}\tb^2\tb^2)}=Z_6(\tilde\a_3,\tilde\a_2,\tilde\a_1)\,.
\ee
Expanding then $Z_6$ in grassmann variables we find how the operators 
of different dimension mix under renormalization:
\be
4\pi \ddt \left(\ba{l}\l_6\\\l_6 \GG\\\l_6 \GG^\dagger\\\l_6{\Z}\ea\right) =
 \g_6\a\matr{~~1~~&~~0~~&~~0~~&~~0~~\\
 			\mg&1&0&0 \\
 			\mg^*&0&1&0 \\
  			2|\mg|^2& \mg^*& \mg&1} 
 \left(\ba{l}\l_6\\\l_6 \GG\\\l_6 \GG^\dagger\\\l_6 {\Z}\ea\right),
 \label{eq:running}
\ee
where $t=\ln(\mu^2)$ and a summation on the different gauge groups is
implicit in the r.h.s..  The gaugino masses $\mg$ and gauge couplings
$\a$ follow the equations $\dot{m}_{\tilde g}/\mg=\dot\a/\a=b\a/4\pi$.

The equations~(\ref{eq:running}) are solved in terms of the evolution
of the gauge coupling constants $\a$ from the GUT to the SUSY scale by
using the auxiliary functions
$$
 R=\frac{\a(S)}{\a(G)}\,,\qquad R_1=\frac{\g_6}{b}(R-1)\,,\qquad R_2=\frac{\g_6}{b}(R^2-1)\,.
$$
The result is:
\bea
\l_6(S)&\!=&\!R^{-\frac{\g_6}{b}}\l_6(G)\0\\[1ex]
%\GG(S)&\!=&\!\GG(G)+\frac{\g_6}{b}\bigg(\frac{\a(S)}{\a(G)}-1\bigg)\mg(G)\0\\
\GG(S)&\!=&\!\GG(G)+R_1\,\mg(G) \0\\[1ex]
{\Z}(S)&\!=&\!{\Z}(G)+2R_1\,\Re[\GG(G)\,\mg^*(G)]+\left(R_2+R_1^2\right)\mg^2(G)\,.\0
\eea 
For example, the largest effect comes from SU(3) color, for which
$\g_6^{(3)}=-4/3$, $b^{(3)}=-3$, $\a_3(M_Z)=0.119$ and $\a_3(G)=1/23$:
\bea
\l_6(S)&\!\simeq&\!1.55\,\l_6(G)\,,\\[1ex]
%\GG(S)&\!=&\!\GG(G)+\frac{\g_6}{b}\bigg(\frac{\a(S)}{\a(G)}-1\bigg)\mg(G)\0\\
\GG(S)&\!\simeq&\!\GG(G)+0.74\,{\mg}_3(G)\,,\\[1ex]
{\Z}(S)&\!\simeq&\! {\Z}(G)+1.48\, \GG(G)\, {\mg}_3(G)+3.26\, {\mg^2}_3(G)\,,
\label{eq:runGUT}
\eea 
where for simplicity we assumed $\GG$ and $\mg$ real.

\pagebreak[3]

\medskip

\pippo{Dressing:}

The effective strength of the dimension 6, 5 and 4 operators can be
compared after dressing with exchange of some gaugino. As shown in
figure~\ref{fig:dressing}, it is clear that the chiral structure of
the D=5 operators requires a Majorana mass to perform a chirality
flip, and for low momentum processes as proton decay this is true also
for the D=4 operators.  The D=5 operators can be dressed by gluino
exchange; the D=4 operators on the other hand can only involve one
gluino exchange while the other loop is necessarily formed via W-ino
(or B-ino exchange).

\begin{figure}[t]
\centerline{\includegraphics[width=25em]{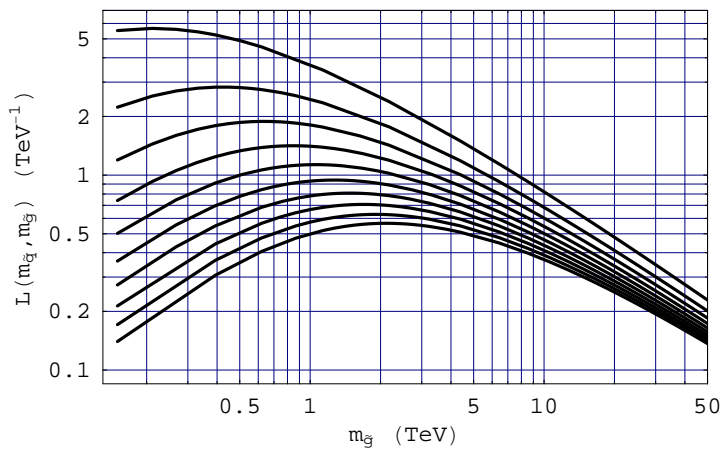}}
\caption{Loop function $L$ in $\TeV^{-1}$, with $m_{\tilde q}$ ranging
 from $0.1$ (upper) to $1\,\TeV$ (lower) in steps of $100\,\GeV$.}
\label{fig:loop}
\end{figure}

We can estimate the strength of the two new operators by defining the 
corresponding effective four-fermion couplings at the SUSY scale:
\bea
%&&\!\!\!  \l_6=\l_6  \0\\[1.6ex]
&&\!\!\!  \l_5=\l_6\,2\,\GG\,
            \frac{\a_3}{4\pi}\,L(m_{\tilde x},m_{\tilde y},{\mg}_3)\0\\[1ex]
&&\!\!\!  \l_4=\l_6\,\,{\Z}\,
            \frac{\a_3}{4\pi}\,L(m_{\tilde x},m_{\tilde y},{\mg}_3)\,
	    \frac{\a_2}{4\pi}\,L(m_{\tilde x'},m_{\tilde y'},{\mg}_2)\0
\eea
where $m_{\tilde x, \tilde y}$ are the sfermion masses entering the
dressing loop(s) and $\a_2$, $\a_3$ the gauge coupling constant
involved. All quantities are evaluated at the SUSY scale. L is the
loop integral:
\bea
L(m_1,m_2,m_3)&=&m_3\frac{
    m_1^2 m_2^2\log\frac{m_1^2}{m_2^2}
   +m_2^2 m_3^2\log\frac{m_2^2}{m_3^2}
   +m_1^2 m_3^2\log\frac{m_3^2}{m_1^2}
     }{
   \left(m_1^2-m_2^2\right) \left(m_1^2-m_3^2\right)
   \left(m_2^2-m_3^2\right)}\0\\
&=&\frac1{m_3}\,\frac{\frac{m^2}{m_3^2}-1-\log \frac{m^2}{m_3^2}}
        {\big(\frac{m^2}{m_3^2}-1\big)^2}\,\qquad\mbox{for }m_1=
        m_2=m
\label{eq:loop}
\eea
The loop integral is plotted in figure~\ref{fig:loop}, where all the
masses are measured in \TeV.  From there we see that $L$ may be of
order $1\,\TeV^{-1}$ (with a maximum of $L\sim5$--$6\,\TeV^{-1}$) for
small sfermion masses $\sim100\,\GeV$. On the other hand the gaugino
mass may be raised up to $1$--$2\,\TeV$ before starting to suppress
the loop.

\pippo{Estimate}

The present limits~\cite{pdg} allow squark masses as low as
$100\,\GeV$ when the gluino mass is $\gtrsim 500\,\GeV$.  In this
parameter region, we find that $L$ is almost maximal, $L\sim 5$, so
for the numerical estimates we will stick to this choice.  A different
choice can be easily considered by extracting the relevant loop factor
from eq.~(\ref{eq:loop}) or directly from figure~\ref{fig:loop}.

With this choice the strengths of the new operators relative to the D=6 one are:
\bea
\l_5/\l_6 &\simeq&  \frac{\a_3}{4\pi}\, 10\, \frac{\GG}{\TeV} 
           \simeq\frac{\GG}{10\,\TeV}\0\\[1ex]
\l_4/\l_6 &\simeq&  \frac{\a_3}{4\pi}\,\frac{\a_2}{4\pi}\,
25\,\frac{{\Z}}{\TeV} \simeq\frac {{\Z}}{(30\, \TeV)^2}\,.
\label{eq:estimate}
\eea
As a result, the effect of D=5 and D=4 operators may be comparable (or
larger) than that of the standard D=6 operators.  However for this to
happen the effective susy-breaking terms $\GG$ and ${\Z}$ should be
larger than the soft masses.  One needs for example $\GG\simeq
10\,\TeV$, a factor of 20 or 100 larger than the gaugino or sfermion
masses.

One should also ask whether these large values might be generated in
the evolution of 13 orders of magnitude from the GUT down to the SUSY
scale, by mixing with other soft parameters, namely the gaugino
masses.  However from the running~(\ref{eq:runGUT}) we see that in the
regime of $\GG, {\Z}\gg {\mg}_3$ the gaugino gives a small
contribution that does not  modify the estimate~(\ref{eq:estimate}).

%In fact
%using~(\ref{eq:runGUT}) to rewrite $\GG$ and ${\Z}$ at GUT scale, we find
%that there is only an additive mixing with ${\mg}_3$:
%%
%\bea
%\GG(S)&=&\GG(G)+0.3\, {\mg}_3\0\\
%{\Z}(S)&=&{\Z}(G)+0.5\,{\mg}_3\,\GG(G)+0.6\,{\mg^2}_3\0
%\eea
%%
%In the regime of $\GG, {\Z}\gg {\mg}_3$ the gaugino gives a small contribution that
%does not substantially modify the estimate~(\ref{eq:estimate}).

Is it then plausible for $\GG$ or ${\Z}$ at GUT scale to be so larger
than other soft susy breaking parameters in the theory?  We argue that
this is possible without spoiling the framework of low energy
supersymmetry.  The reason is as follows:
from the expression of $\GG$ and ${\Z}$, eq.~(\ref{eq:effectivesoftterms}), 
we see that they are induced, in addition to the soft masses of fermion fields, by $m^2_H$
and $f_H$, the soft SUSY-breaking parameters in the heavy
higgs sector.  One can not play much with the soft masses of the heavy
fields $m^2_H$, since these are constrained because they usually mix
with the MSSM higgses soft masses in the renormalization from the
Planck to the GUT scale.  On the other hand the F-terms $f_H$ are less 
constrained, since they do not directly enter in the running and one
should not assume them to be small.  This can be seen in the minimal
SU(5) model as we illustrate in the following section.

\section{SU(5) example}

\pippo{SU(5):   $\l_6$, $\GG$, ${\Z}$ in terms of GUT parameters. }

In minimal supersymmetric SU(5)~\cite{su5} the GUT breaking is due to
the VEV of an adjoint superfield $\S\in\R{24}$. The superpotential for
$\S$ includes the soft terms $A_\S$, $B_\S$ as follows:
\be
W(\S)=M_\S\tr\S^2 (1-B_\S\t^2)+\frac16\l_\S\tr\S^3 (1-A_\S\t^2)\,.
\label{eq:WS}
\ee

The effect of $A_\S$ and $B_\S$ is to give an F-term to
$\langle\S\rangle$ and to shift its magnitude by a small amount:
\be
\label{eq:sigma}
\langle\S\rangle=v_\S\bigg[1+(A_\S-B_\S)\t^2\bigg]\l_{Y}\,,\qquad
v_\S=8\sqrt{15}\frac{M_\S}{\l_\S}\bigg[1+\frac{A_\S-B_\S}{2M_\S}\bigg]\simeq8\sqrt{15}\frac{M_\S}{\l_\S}\,,
\ee
where $\l_{Y}=\mbox{diag}(2,2,2,-3,-3)/\sqrt{60}$. 

The fermion multiplets in SU(5) are $\R{10}$ and $\Rb5$, and proton
decay can proceed via the four field operators involving the
combinations $\R{10}$-$\R{10}$-$\Rb{5}$-$\Rb{5}$ or
$\R{10}$-$\R{10}$-$\R{10}$-$\R{10}$.  At GUT scale the standard D=6
operator has coupling constant
\be
\l_6(G)={g_5^2}/{M_A^2}\,,\qquad \mbox{with }M_A^2=5\,g_5^2 v_\S^2/12\,,
\ee
where $g_5$ is the SU(5) gauge coupling.  Using
eq.~(\ref{eq:effectivesoftterms}) we find the coefficients of the new
operators:
\vspace*{-1ex}
\bea
\GG=B_\S-A_\S\,,\qquad
{\Z}_{\R{10}\,\Rb5}&=&-m^2_{\R{10}}-m^2_{\Rb5}+m^2_{\S}+|B_\S-A_\S|^2\0\\[1ex]
{\Z}_{\R{10}\,\R{10}}&=&-2m^2_{\R{10}}+m^2_{\S}+|B_\S-A_\S|^2\,,
\label{eq:A-B}
\eea
where $m^2_{\Rb5}$, $m^2_{\R{10}}$ and $m^2_\S$ are the soft masses of
the $\Rb5$, $\R{10}$ fermion multiplets and of $\S$ itself.

In the previous section we found that the new operators for proton
decay are relevant when the squark masses are small while $\GG$ or
${\Z}$ are larger, $\sim 10\,\TeV$.  From~(\ref{eq:A-B}) we see
that this may be realized when the soft mass $m_\S^2$ or the analytic
soft terms $B_\S-A_\S$ are large.  A large $m_\S^2$ is not appealing,
since $m_\S$ enters the RG running of the Higgs soft masses and would
induce large values for these, spoiling the picture of electroweak
breaking. The same holds for $A_\S$, since it also enters in the
running of $m_\S^2$ and other soft parameters (see
e.g.~\cite{polonski}), and a large $A_\S$ would indirectly cause color
breaking minima.  On the other hand we note that $B_\S$ does not enter
the evolution of other quantities and may be sensibly large, without
driving all the other soft parameters to large values as well.

\medskip 
\pippo{Running from Planck to GUT:}

The fact that soft $B$-terms do not enter in any beta function is a
general statement valid in the MS scheme, that follows from SUSY and
the fact that in this scheme no spurious scales are introduced.  It
turns out that the soft $B$-terms like $B_\S$, being of dimension one,
never enter any RG equation, at all loops.  Specifically, this can be
verified in RG equations of soft masses, for $A$-terms and for
$B$-terms themselves; for example we have, for the one-loop running of
$A_\S$ and $B_\S$ from Planck to GUT scale:%
\footnote{The constants $\l_H$, $A_H$ are defined
  below,~(\ref{eq:WH}).  To compare with evolution of other quantities
  see e.g.~\cite{polonski}.}
\bea
16\pi^2 \ddt A_\S&=&\frac{63}{20} A_\S\l_\S^2+3A_H\l_H^2-30\,g_5^2 \,{\mg}_5\\
16\pi^2 \ddt B_\S&=&\frac{21}{10} A_\S\l_\S^2+2A_H\l_H^2-20\,g_5^2 \,{\mg}_5\,.
\eea
%
%(ESTRAPOLAZIONE DA POLONSKY con 2/3)
From these equations one can also see that if a large $B_\S$ is
generated at the Planck scale, it will not be substantially affected
running down to the GUT scale.

Of course one can not hope to raise one soft parameter without
consequences.  Even if $B_\S$ does not appear in RG equations, going
from the MS to a physical scheme, it will enter in finite corrections.
In particular a large $B_\S$ will generate corrections to soft
quantities~\cite{hisano,polonski}, raising them as if actually SUSY
were broken at the $B_\S$ scale.  As we discuss below, this effect is
relevant only when $B_\S$ exceeds $\sim50\,\TeV$, with some model
dependence. Below this limit the only physical effect of a large
$B_\S$ is in the soft $\GG$ and ${\Z}$ coefficients, where it can
directly dominate in the D=5 and D=4 operators and thus enhance or
even suppress the proton decay rate.

\medskip

To give a concrete estimate in the SU(5) example, we assume large
$B_\S$ and use the soft coefficients $\GG\simeq B_\S$ and $\Z\simeq
|B_\S|^2$ in eq.~(\ref{eq:estimate}), to find how the proton lifetime
for a gauge-mediated channel like $p\to\pi^0 e^+$ is modified:
\be
\tau_{SOFT}^{-1}\simeq\tau^{-1}_{SUSY}\left(1 + \frac{B_\S}{10\,\TeV} + 0.1\left| \frac{B_\S}{10\,\TeV}\right|^2\right)^2.
\ee

\begin{figure}[t]
\centerline{\includegraphics[width=25em]{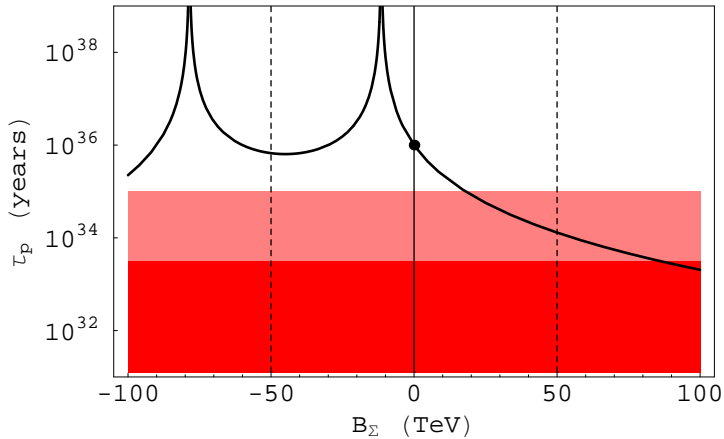}}
\vspace*{-1ex}
\caption{Effect of dimension 4 and 5 on gauge mediated proton decay as
  a function of the soft SUSY-breaking terms in the heavy sector.  The
  dot represents our reference value for the supersymmetric gauge
  mediated proton lifetime, taken to be $10^{36}\,{\rm y}$.  The
  shaded regions show current and 10-years expected limits on the
  $p\to\pi^0 e^+$ partial lifetime. Dashed lines mark the limit
  where $B_\S$ start to affect the higgs soft masses and other soft
  parameters.}
\label{fig:rate}
\end{figure}

An explicit plot of the effect of large $B_\S$ is shown in
figure~\ref{fig:rate}, where we assume a reference value of
$10^{36}\,\mbox{y}$ for the proton partial lifetime.\footnote{This
  reference value corresponds to a Grand Unification scale of
  $2\cdot10^{16}\,\GeV$, and we remind that $\tau^{-1}_{SUSY}$ scales
  as the fourth power of $\MG$, which is model dependent.}
We see that the effect can be rather evident: for example for negative
$B_\S$ the proton decay can be made absolutely unobservable,
while for $B_\S$ positive one can enter in the region of
sensitivity of the next ten years water-cerenkov
detectors~\cite{HyperK}.  We conclude that large soft SUSY-breaking
terms for the heavy fields may significantly affect the proton decay
rate even in the gauge mediated channels.

\medskip

Let us now address the fine tuning problems related to the stability
of the hierarchy in presence of a large $B_\S$.  This point is related
to the problem of doublet-triplet splitting, as can be seen in the
SU(5) example.

% This point is related to the problem of doublet-triplet splitting, And
% goes beyond the scope of this paper, since it would involve the choice
% of specific models.
% For instance, in models where the Higgs comes as a pseudo-goldstone
% boson of some global symmetry~\cite{higgs-pgb} this works partially:
% without arranging any structure in the soft sector the higgs mass
% matrix has large entries and the radiative violation of the global
% symmetry can lead to the effective nonzero higgs mass that is larger
% than desired.

The superpotential involves also the higgses $H$, $\overline H$
(transforming in the \R{5}, \Rb{5}) 
\be
W(\S,H,\bar{H})=W(\S)+ M_H(1-B_H\t^2)\bar{H}H + \l_H(1-A_H\t^2) \bar{H}\S H  \,.
\label{eq:WH}
\ee
where $W(\S)$ is given in~(\ref{eq:WS}). After the SU(5) breaking
$H$, $\overline H$ leave the light MSSM higgs doublets $H_u$, $H_d$,
with their soft masses $m_u^2$, $m_d^2$, and we get also the effective $\mu$
and $B_\mu$ terms
\be
\mu (1-B_\mu\t^2)H_uH_d
\ee
where
\bea
\label{eq:mu}
\mu &=& M_H - \frac{3}{\sqrt{60}} \l_H v_\S \, ,\\
\label{eq:Bmu}
\mu B_\mu &=& \frac{3}{\sqrt{60}}\l_H v_\S (A_\S-B_\S-A_H+B_H) +O(B_\S-A_\S)^2\,.
\eea
Therefore two fine-tuning conditions are needed to achieve the
electroweak symmetry breaking at the correct scale, one for $\mu$ and
another for $B_\mu$. In fact the mass matrix of the higgs scalars is: 
\be
\label{eq:Hmatrix}
\matr{\mu^2+m^2_u&\mu B_\mu\\
       \mu B_\mu&\mu^2+m^2_d}
\ee
and all entries should be of the order of the electroweak scale. 

The second fine tuning~(\ref{eq:Bmu}) can be avoided by assuming
universality of $A$ and $B$ terms separately, $A_\S = A_H$ and $B_\S =
B_H$, as noticed in~\cite{kmy}, so that $(A_\S-B_\S-A_H+B_H)= 0$ and
$B_\mu$ is of the order of soft susy-breaking scale.  In the case of
large $B_\S\sim10\,\TeV$ however one still gets a $B_\mu$ term that is
too large, therefore the right pattern of electroweak breaking can be
obtained only by tuning the two independent parameters, the
supersymmetric $\mu$ and soft $B_\mu$.

Of course, this minimal SU(5) model is not realistic, and one should
not be surprised to find that fine tunings are required. In the next
section we describe how in specific models fine tunings can be avoided
and one can have large $B$-terms without spoiling the hierarchy.

\medskip

Before moving to more realistic models, we point out that generically
there are also finite corrections induced by $B$-terms. For example in
SU(5) $B_\S$ induces a shift of the analytic soft term for the higgses
$B_\mu$, of their soft masses and also of the gaugino masses.  These
corrections are loop suppressed:
\be
\label{eq:finite}
\delta B_\mu\sim\frac{\l_H^2}{(4\pi)^2}B_\S\,,\qquad
\delta \mg\sim\frac{g_5^2}{(4\pi)^2}B_\S\,,
\ee
with some model dependent numerical factors~\cite{thrash-old}. Since
$\l_H\simeq g_5\simeq0.7$, the loop suppression factor is $\sim1/100$,
and we conclude that these corrections can be ignored as far as
$B_\S<50\,\TeV$.  Beyond this limit the gaugino mass and the higgs
mass terms would need some fine tuning, to avoid breaking SUSY at a
high effective scale or having an unacceptably large higgs mass.

% As we pointed out before, the $A$'s cannot be taken very large, since
% large $A$'s for Yukawas would generate color-breaking vacuum.
% Therefore one is led to consider large $B_\S$ only. In this case the
% second term in (\ref{eq:Bmu}) gets too large, of the order of the
% geometrical mean between $B$ and the electroweak scale.

\section{Realistic models} 

In minimal SU(5) model the problem of doublet-triplet splitting has
only a technical solution: fine tuning of $\mu$, eq.~(\ref{eq:mu})
that is stable against radiative corrections; the situation is then
worsened by the need of another fine tuning in the soft
terms, eq.~(\ref{eq:Bmu}).

Moreover, to achieve the right electroweak scale of order $100\,\GeV$
with a large $B_\mu\sim10\,\TeV$ would require a fine tuning with the
$\mu$ term, while there is no apriori correlation between these two
parameters.

This problem gets another twist in realistic models in which the
doublet-triplet splitting problem is solved without fine tuning. In
particular, in SU(5) this can be done via the "Missing Doublet
Mechanism" (MDM) \cite{MDM}, and in SO(10) via the "Missing VEV
Mechanism" (MDM) \cite{MVM}, while in SU(6) via the "Goldstones
instead of Fine Tuning" (GIFT) Mechanism \cite{GIFT}.

In particular, in all these models the soft parameters like $B_\S$ or
$A_\S$ for the heavy GUT breaking superfields can be taken much larger
than that of matter superfields, without creating additional
fine-tuning problems.  Let us briefly describe them here.

\medskip

In SU(5), the missing doublet model \cite{MDM} contains the Higgs
superfields in representations $\Phi \sim 75$, $H\sim 5$, $\bar H \sim
\bar 5$, $\Psi \sim 50$, $\bar \Psi \sim \overline{50}$, with the
following superpotential terms:
\be
W = M\Phi^2 + \l \Phi^3 + M_1 \Psi \bar{\Psi} + \l_1 H \Phi \bar{\Psi} + 
\l_2 \bar{H} \Phi \Psi + \mu H\bar{H} 
\ee 
with $M$ and $M_1$ being the mass parameters order $M_G$ and $\l$'s
being the order 1 coupling constants. SU(5) is broken to
SU(3)$\times$SU(2)$\times$U(1) by the VEV of $\Phi$ which also
generates the mixing between the color triplet fragments in the Higgs
5- and 50-plets, whereas there are no doublets in the 50-plets.  In
this way, all color triplets are heavy, with mass of order $M_G$,
while the doublets in $H,\bar H$ remain light, with mass given by the
$\mu$-term.  Obviously, in this theory the soft parameter $B_\Phi$ can
be taken large without inducing a large $B_\mu$  (still inside the
limits set by the induced finite corrections like~(\ref{eq:finite})).

\medskip

For SO(10), in the missing VEV model \cite{MVM}, the philosophy is
similar: the Higgs doublets remain massless because the GUT-breaking
fields have zero VEV along the direction that would give them a mass,
whereas it couples to the triplets with non-zero VEV.  Therefore also
in this case the protection of the doublet sector is due to group
theoretical reasons, therefore large soft terms $\sim10\,\TeV$ in the
heavy sector will not influence $\mu$ and $B_\mu$ and the electroweak
scale will not be destabilized.

\medskip

In the SU(6) model~\cite{GIFT}, the SU(6) gauge symmetry is broken by
two sets of superfields: one contains an adjoint representation $\S
\sim 35$, that leads to the breaking channel SU(6)$\to$SU(4)$\times$
SU(2)$\times$ U(1), and the other contains two fundamental
representations $H\sim 6$ and $\bar H \sim \bar 6$ that break
SU(6)$\to$SU(5). As a result the two channels together break the SU(6)
gauge symmetry down to SU(3)$\times$SU(2)$\times$U(1). Also, one
assumes that the Higgs superpotential does not contain the mixed term
$H \S \bar H$, so that it has the form $W = W(\S) + W(H,\bar H)$,
where
\be
W(\S) = M\S^2 + \l \S^3\,,\qquad W(H,\bar H) = Y (H \bar H - V^2)\,.
\ee
As a result there is an accidental global symmetry SU(6)$_\S
\times$SU(6)$_H$, which independently transform $\S$ and $H,\bar H$
superfields. Then, in the limit of unbroken supersymmetry the MSSM
Higgs doublet $H_u$, $H_d$ appear as massless goldstone superfields
built up as a combination of doublet fragments from $\S$ and $H$,
$\bar H$, that remain uneaten by the gauge bosons. Therefore in this
limit $\mu$ vanishes exactly.

Supersymmetry breaking terms like $A_\S$, $B_\S$ shift the VEVs and
also give F-terms to them, therefore generating $B_\mu$ term for the
MSSM Higgses. However, since these terms also respect the global
symmetry SU(6)$_\S\times$SU(6)$_H$, the mass matrix of the
Higgses~(\ref{eq:Hmatrix}) is degenerate and so one Higgs scalar
(combination of the scalar components of $H_u$ and $H_d$) still
remains massless.  Thus, even with arbitrary $B_\S$ that give $\mu\sim
B_\mu\sim B_\S$, there is an automatic relation between $\mu$ and
$B_\mu$ terms that guarantees that the determinant
of~(\ref{eq:Hmatrix}) vanishes.

This degeneracy is removed only by radiative corrections due to Yukawa
terms that do not respect the global symmetry, and the resulting Higgs
mass will be of the order of $\mu$ and $B_\mu$, given by the mismatch
in their renormalization.  Therefore, in the case of large $B_\S\sim
10\,\TeV$ we are still left with a ``little'' hierarchy problem of the
electroweak scale stability against $10\,\TeV$. However by enlarging
the gauge symmetry this issue can be avoided. In fact one can have
that 10$\,\TeV$ is only an intermediate scale where an extra global
symmetry guarantees the protection of the electroweak scale, the so
called super-little-higgs mechanism~\cite{higgs-pgb}.

\section{Conclusions}

In this paper we have studied the effects of soft SUSY-breaking terms
on proton decay in SUSY GUT theories.  While the dominant effect in
SUSY GUT comes from D=5 higgs-mediated operators, these are very model
dependent and may be suppressed by specific constructions.  Here we
have focused on gauge mediated effective operators, that are usually
unavoidable.

We have shown how soft terms enter into the gauge-mediated effective
operators for proton decay: while the supersymmetric operators are of
dimension 6, SUSY breaking always induces new operators of dimension 5
and 4.

We computed their renormalization from the GUT to the SUSY scale, that
amounts to a small mixing of the D=6, 5, 4 operators through the
gaugino masses.

The new operators are dressed via gaugino exchange and transformed
into D=6 four-fermion operators, and have the same suppression factor
$M_G^{-2}$ of the standard D=6 operators.  They however have numeric
coefficients that depend on the ratio of soft-breaking parameters in
the heavy and light sectors.

When all the soft breaking parameters are of the same order, the
dressing loop factors are small enough to suppress these new
operators.  However, we note that the $B$-terms in the heavy-Higgs
sector may be substantially higher than the standard soft masses, and
they do not mix with soft masses under renormalization.  Finite
corrections are present which are irrelevant when the heavy
$B$-terms are smaller than $\sim50\,\TeV$.

The heavy higgses soft-terms then enter the GUT breaking process and lead
to observable effects on D=6 proton decay. $B$-terms as low as
$10\,\TeV$ can lead to substantial effects on the proton decay and,
depending on their sign, may enhance or even suppress the proton decay
rate in gauge mediated channels.

\section{Acknowledgments}
We would like to thank F. Vissani for useful discussions about current
and future experimental limits on proton lifetime.  This work was
partially supported by the MIUR grant under the Projects of National
Interest PRIN 2004 ``Astroparticle Physics''.


\begin{thebibliography}{99}

\bibitem{gut}
  For a general review see P.~Langacker,
  % ``Grand Unified Theories And Proton Decay,''
  %
  \emph{Phys.\ Rept.}  {\bf 72} (1981) 185. 
  %%CITATION = PRPLC,72,185;%%


\bibitem{nathreview} 
    P. Nath, P.F. Perez, \emph{"Proton stability in
    grand unified theories, in strings, and in branes"}, {\tt
    hep-ph/0601023}.

\bibitem{W} S. Weinberg, \emph{Phys.\ Rev.} {\bf 43} (1979) 1566;\\
            F. Wilczek and A. Zee, \emph{Phys.\ Rev.} {\bf 43} (1979) 1571.

\bibitem{john}
  Z.G.~Berezhiani and J.L.~Chkareuli,
  %``Proton Decay In Grand Unified Models With Horizontal Symmetry,''
  \emph{JETP Lett.}  {\bf 38} (1983) 33
  [\emph{Pisma Zh.\ Eksp.\ Teor.\ Fiz.}  {\bf 38} (1983) 28].
  %%CITATION = JTPLA,38,33;%%
  %``Quark - Leptonic Families In A Model With SU(5) X SU(3) Symmetry.''
 \emph{ Sov.\ J.\ Nucl.\ Phys.} {\bf 37} (1983) 618
  [\emph{Yad.\ Fiz.}  {\bf 37} (1983) 1043]; 
  %%CITATION = SJNCA,37,618;%%


\bibitem{dimfive}
  S.~Weinberg,
  %``Supersymmetry At Ordinary Energies. 1. Masses And Conservation Laws,''
 \emph{Phys.\ Rev.} {\bf D 26} (1982) 287;\\
  %%CITATION = PHRVA,D26,287;%%
  N.~Sakai, T.~Yanagida,
  %``Proton Decay In A Class Of Supersymmetric Grand Unified Models,''
  \emph{Nucl.\ Phys.} {\bf B 197} (1982) 533.
  %%CITATION = NUPHA,B197,533;%%


\bibitem{bbv}
  G.D.~Coughlan et al,
   %``Baryogenesis, Proton Decay And Fermion Masses In Supergravity Guts,''
   %
   \emph{Phys.\ Lett.} {\bf B158} (1985) 401;\\
  %%CITATION = PHLTA,B158,401;%%
  %%CITATION = HEP-PH 9809301;%%
  K.S. Babu, S.M. Barr, 
  \emph{Phys.\ Rev.} \textbf{D48}  (1993) 5354--5364 
  {\tt [hep-ph/9306242]};\\
  R.~Barbieri et al., 
  %  G.R.~Dvali, A.~Strumia, Z.~Berezhiani and L.J.~Hall,
  %  ``Flavor in supersymmetric grand unification: A Democratic approach,''
 \emph{Nucl.\ Phys.} {\bf B432} (1994) 49   {\tt [hep-ph/9405428]}; \\ 
  %%CITATION = HEP-PH 9405428;%%
  
\bibitem{bvt}
   Z.~Berezhiani,
  ``Fermion masses and mixing in SUSY GUT,''
  {\tt hep-ph/9602325};\\
  %%CITATION = HEP-PH 9602325;%%
  Z.~Berezhiani, Z.~Tavartkiladze and M.~Vysotsky,
  ``d = 5 operators in SUSY GUT: Fermion masses versus proton decay,''
  {\tt hep-ph/9809301};\\
  Z.G. Berezhiani, 
  %\emph{Predictive SUSY ${\rm SO}(10)$ model with very low $\tan\beta$},
  \plb{355}{1995}{178} [\hepph{9505384}];\\
  Z.~Berezhiani and F.~Nesti,
   %``Supersymmetric SO(10) for fermion masses and mixings: Rank-1 structures  of flavour,''
  JHEP {\bf 0603} (2006) 041
  {\tt [hep-ph/0510011]}.
  
\bibitem{gia}
  G.R.~Dvali,
  %``Light Colour-Triplet Higgs is Compatible with Proton Stability: An
  %alternative approach to the doublet-triplet splitting problem,''
  %
  \emph{Phys.\ Lett.} {\bf B372} (1996) 113  {\tt [hep-ph/9511237]}.
  %%CITATION = HEP-PH 9511237;%% 


\bibitem{Rparity}
  A.~Salam and J.A.~Strathdee,
  %``Supersymmetry And Fermion Number Conservation,''
  \emph{Nucl.\ Phys.}  {\bf B87} (1975) 85;\\
  %%CITATION = NUPHA,B87,85;%%
  P.~Fayet,
  % ``Supergauge Invariant Extension Of The Higgs Mechanism And A Model For The
  %Electron And Its Neutrino,''
  \emph{Nucl.\ Phys.}  {\bf B90} (1975) 104.
  %%CITATION = NUPHA,B90,104;%%

%\bibitem{Hall:1983iz}
%  L.J.~Hall, J.D.~Lykken and S.~Weinberg,
%  %``Supergravity As The Messenger Of Supersymmetry Breaking,''
%  \emph{Phys.\ Rev.} {\bf  D 27} (1983) 2359.
%  %%CITATION = PHRVA,D27,2359;%%


\bibitem{DS}
 J.P. Derendinger and C. Savoy \emph{Phys.\ Lett.} {\bf B118} (1982) 347.

\bibitem{sak1}
 N. Sakai \emph{Phys.\ Lett.} {\bf B121} (1983) 130.

\bibitem{sak2}
 N. Sakai, \emph{Nucl.\ Phys.} {\bf B238} (1984) 317;\\
 N. Haba, N. Okada, {\tt hep-ph/0601003}.

\bibitem{decoupling}
  A.~Pomarol and S.~Dimopoulos,
  %``Superfield derivation of the low-energy effective theory of softly broken
  %supersymmetry,''
  \emph{Nucl.\ Phys.} {\bf B453} (1995) 83
  {\tt [hep-ph/9505302]};\\
  R.~Rattazzi,
  %``A Note on the effective soft SUSY breaking Lagrangian below the GUT
  %scale,''
  \emph{Phys.\ Lett.} {\bf B375}  (1996) 181
 {\tt [hep-ph/9507315]}.
  %%CITATION = HEP-PH 9507315;%%

\bibitem{fayet}
  P. Fayet, \emph{Nuovo Cim.} {\bf 31 A} (1976) 327.

%\bibitem{new}
%  N. Haba, N. Okada, {\tt hep-ph/0601003}

% \bibitem{gato}
%   B. Gato, J. Leon, J. Perez-Mercader and M. Quiros
%   %``Renormalization group analysis for a general softly broken
%   %supersymemtric gauge theory''
%  \emph{Nucl.\ Phys.}  {\bf B 253} (1985) 285 {\tt [hep-ph/????]}.

\bibitem{kazakov}
  L.V. Avdeev, D.I. Kazakov, I.N. Kondrashuk,
  % ``Renormalization in softly broken SUSY gauge theory'',
  %``Renormalizations in softly broken SUSY gauge theories,''
  \emph{Nucl.\ Phys.}  {\bf B510} (1998) 289 {\tt [hep-ph/9709397]}. 
  %%CITATION = HEP-PH 9709397;%%
  See also G.F.~Giudice and R.~Rattazzi,
  % ``Extracting supersymmetry-breaking effects from wave-function
  % renormalization,''
  %
  \emph{Nucl.\ Phys.} {\bf B511} (1998) 25
  {\tt [hep-ph/9706540]}.\\
  %%CITATION = HEP-PH 9706540;%%

\bibitem{ellis}
  A.J.~Buras, J.R.~Ellis, M.K.~Gaillard and D.V.~Nanopoulos,
  %``Aspects Of The Grand Unification Of Strong, Weak And Electromagnetic
  %Interactions,''
  \emph{Nucl.\ Phys.} {\bf B135} (1978) 66.
  %%CITATION = NUPHA,B135,66;%%

\bibitem{ibanez}
  L.E. Ibanez, C. Munoz, \emph{Nucl.\ Phys.} {\bf B245} (1984) 425.

\bibitem{su5}
  S. Dimopoulos and H. Georgi, \emph{Nucl.\ Phys.} {\bf B193} (1981)  150;\\
  N. Sakai, \emph{Z. Phys.} {\bf C11}  (1981) 153. 

\bibitem{hisano}
  J.~Hisano, H.~Murayama and T.~Goto,
  %  ``Threshold correction on gaugino masses at grand unification scale,''  
  \emph{Phys.\ Rev.} {\bf D49} (1994) 1446.  
  %%CITATION = PHRVA,D49,1446;%%

\bibitem{polonski}
  N.~Polonsky and A.~Pomarol,
  %``Nonuniversal GUT corrections to the soft terms and their implications in
  %supergravity models,''
  \emph{Phys.\ Rev.}  {\bf D51} (1995) 6532
  {\tt [hep-ph/9410231]}.

\bibitem{HyperK}
  Super-Kamiokande Collaboration, \emph{Phys.\ Rev.\ Lett.} {\bf 81} (1998) 3319.
  The future sensitivity of current and proposed water-cerenkov
  detectors like HyperK or Uno is somehow difficult to extract from
  current literature.  With some degree
  of optimism we believe that a limit of $10^{35}\,$y in ten years is
  a realistic goal.  See also~\cite{nathreview}.

%(CITE RUBBIA??  {\tt [hep-ph/0407297]})
%
%(CITE UNO??  {\tt [hep-ph/0005046]})

\bibitem{thrash-old}
 J.~Hisano, H.~Murayama and T.~Goto,
  %``Threshold correction on gaugino masses at grand unification scale,''
  \emph{Phys.\ Rev.} {\bf D49}, 1446 (1994).
  %%CITATION = PHRVA,D49,1446;%%

\bibitem{kmy}
  Y.~Kawamura, H.~Murayama and M.~Yamaguchi,
  % ``Low-energy effective Lagrangian in unified theories with nonuniversal
  % supersymmetry breaking terms,''
  %
  \emph{Phys.\ Rev.} {\bf D51}, 1337 (1995)
  [hep-ph/9406245].
  %%CITATION = HEP-PH 9406245;%%


\bibitem{MDM} 
H. Georgi, \emph{Phys. Lett.} {\bf B 108} (1982) 283; \\
A. Masiero, D.V. Nanopoulos, K. Tamvakis and T. Yanagida,  
\emph{Phys. Lett.} {\bf B 115} (1982) 380; \\
B. Grinstein,  \emph{Nucl. Phys.} {\bf B 106} (1982) 387; \\ 
For more details, see also 
J.L. Lopez and D.V. Nanopoulos,  \emph{Phys. Rev.} {\bf D 53} (1996) 2670; \\
Z. Berezhiani and Z. Tavartkiladze,  \emph{Phys. Lett.} {\bf B 396} (1997) 150. 

\bibitem{MVM} 
S. Dimopoulos and F. Wilczek, NSF-ITP-82-07 (unpublished); \\
M. Srednicki,  \emph{Nucl. Phys.} {\bf B 202} (1982) 327; \\
For more details, see also 
K.S. Babu  and S.M. Barr,  \emph{Phys. Rev.} {\bf D 48} (1993) 5354; 
\emph{ibid.} {\bf D 50} (1994) 3529; \\ 
 Z. Berezhiani and Z. Tavartkiladze,  \emph{Phys. Lett.} {\bf B 409} (1997) 220; \\
 Z. Berezhiani and A. Rossi,  \emph{Nucl. Phys.} {\bf B 594} (2001) 113.  

\bibitem{GIFT}
%\cite{Berezhiani:1989bd}  \bibitem{Berezhiani:1989bd}
  Z.G.~Berezhiani and G.R.~Dvali,
  % ``Possible Solution Of The Hierarchy Problem In Supersymmetrical Grand Unification Theories,''
  \emph{Bull.\ Lebedev Phys.\ Inst.}  {\bf 5} (1989) 55
  [\emph{Kratk.\ Soobshch.\ Fiz.} {\bf 5} (1989) 42]; \\
  %%CITATION = SPLRD,5,55;%%
%\cite{Barbieri:1994kw}  \bibitem{Barbieri:1994kw}
  %
%\cite{Barbieri:1993wz}  \bibitem{Barbieri:1993wz}
  R.~Barbieri, G.R.~Dvali and M.~Moretti,
  % ``Back to the doublet - triplet splitting problem,''
  \emph{Phys.\ Lett.} {\bf B312} (1993) 137; \\ 
  %%CITATION = PHLTA,B312,137;%%
  %
  R.~Barbieri et al., 
%  G.~R.~Dvali, A.~Strumia, Z.~Berezhiani and L.~J.~Hall,
 %  ``Flavor in supersymmetric grand unification: A Democratic approach,''
 \emph{Nucl.\ Phys.} {\bf B432} (1994) 49   {\tt [hep-ph/9405428]}; \\ 
  %%CITATION = HEP-PH 9405428;%%
%\cite{Berezhiani:1995sb}  \bibitem{Berezhiani:1995sb}
  %
  Z.~Berezhiani, C.~Csaki and L.~Randall,
  %``Could the supersymmetric Higgs particles naturally be pseudoGoldstone bosons?,''
  \emph{Nucl.\ Phys.}  {\bf B444} (1995) 61 {\tt[hep-ph/9501336]}; \\ 
  %%CITATION = HEP-PH 9501336;%%
%\cite{Berezhiani:1995dt}  \bibitem{Berezhiani:1995dt}
  %
  Z.~Berezhiani,
 %  ``SUSY SU(6) GIFT for doublet-triplet splitting and fermion masses,''
   \emph{Phys.\ Lett.} {\bf B355} (1995) 481  {\tt [hep-ph/9503366]}; 
  %%CITATION = HEP-PH 9503366;%%
%\cite{Berezhiani:1997as}  \bibitem{Berezhiani:1997as}
  %
   %\cite{Berezhiani:1994if}  \bibitem{Berezhiani:1994if}
  %Z.~Berezhiani,
  % ``Solving SUSY GUT problems: Gauge hierarchy and fermion masses,''
  {\tt hep-ph/9412372}; \\ 
  %%CITATION = HEP-PH 9412372;%%
%
%\cite{Dvali:1996sr} \bibitem{Dvali:1996sr}
  G.R.~Dvali and S.~Pokorski,
  % ``Role of the anomalous U(1)A for the solution of the doublet-triplet
  % splitting problem via the pseudo-Goldstone mechanism,''
  \emph{Phys.\ Rev.\ Lett.}  {\bf 78} (1997) 807
  {\tt [hep-ph/9610431]}; \\ 
  %%CITATION = HEP-PH 9610431;%%
  %
  Z.~Berezhiani,
   ``2/3 splitting in SUSY GUT: Higgs as Goldstone boson,''
  {\tt hep-ph/9703426}.
  %%CITATION = HEP-PH 9703426;%%



%\bibitem{BBO} Barger, Berger, Ohmann, Phys.\
%Rev.\ {\bf D47} (1993) 1093--1113 [{\tt hep-ph/9209232}]

\bibitem{pdg} See page 1019 of Particle Data Group, \emph{Phys.\ Lett.} {\bf B592} (2004).


 \bibitem{higgs-pgb}
  Z.~Berezhiani, P.H.~Chankowski, A.~Falkowski and S.~Pokorski,
   %``Double protection of the Higgs potential,''
   \emph{Phys.\ Rev.\ Lett.}  {\bf 96} (2006) 031801
   {\tt [hep-ph/0509311]};\\
   %%CITATION = HEP-PH 0509311;%%
   T.~Roy and M.~Schmaltz,
   %``Naturally heavy superpartners and a little Higgs,''
   JHEP {\bf 0601} (2006) 149
   {\tt [hep-ph/0509357]};\\
   %%CITATION = HEP-PH 0509357;%%
   C.~Csaki, G.~Marandella, Y.~Shirman and A.~Strumia,
   %``The super-little Higgs,''
   \emph{Phys.\ Rev.} {\bf D73} (2006) 035006
   {\tt [hep-ph/0510294]}.
   %%CITATION = HEP-PH 0510294;%%

\end{thebibliography}
\end{document}